\documentclass{PoS}

\title{Evidence for Two-Component Jet in Sw J1644+57}

\ShortTitle{Evidence for Two-Component Jet in Sw J1644+57}

\author{Jiuzhou Wang$^a$,  \speaker{Weihua Lei}$^a$ \thanks{This work is supported by National Basic Research Program (``973'' Program) of China under Grant No. 2014CB845800, National Natural Science Foundation of China under grants 11361140349 (China-Israel jointed program), 11173011 and U1431124, NSF under Grant No. AST-0908362. }, Dingxiong Wang$^a$, Wei Xie$^a$ and Bing Zhang$^b$ \\
\llap{$^a$}
       School of Physics, Huazhong Univ. of Sci. \& Tech.\\
       E-mail: \email{leiwh@hust.edu.cn}  \\
\llap{$^b$}
Department of Physics and Astronomy, University of Nevada Las Vegas
       }

\abstract{The continued observations of Sw J1644+57 in X-ray and radio bands accumulated a rich data set to study the relativistic jet launched in this tidal disruption event. We find that the re-brightening feature in the radio light curve can be naturally explained by the two-component jet model. The possible origin of this structured jet are the Blandford-Znajek and Blandford-Payne mechanisms. We also show that this two-component jet model can interpret the two kinds of quasi-periodic variations in the X-ray light curve: a 200 second quasi-periodic oscillation (QPO) and a 2.7-day quasi-periodic variation. The latter is interpreted by a precessing outer jet launched near the Bardeen-Petterson radius of a warped disk. The $\sim$ 200s QPO could be associated with a second, narrower jet sweeping the observer line-of-sight periodically, which is launched from a spinning black hole in the misaligned direction with respect to the black hole's angular momentum. }

\FullConference{Swift: 10 Years of Discovery\\
		 2-5 December 2014\\
		 La Sapienza University, Rome, Italy}

\begin{document}
\section{Introduction}
Recently, much attention has been paid on the discovery of the hard X-ray transient event
Swift J16449.3+573451, (hereafter ``Sw J1644+57'', \cite{bkg11}). This event has been
interpreted as a tidal disruption event (TDE) with jet \cite{bgm11, ltc11, bkg11, zbs11}.
Its rich observations in $\gamma$-ray, X-ray, radio, mm, infrared bands provide us a good opportunity to study the underlying physics of launching a relativistic jet/outflow from TDE events. Detailed data also allow us to diagnose the composition and the structure of the jet.

The unusual features of Sw J1644+57 in its super-Eddington X-ray luminosity \cite{bkg11},
bright radio afterglow \cite{zbs11}, and a historical stringent X-ray flux upper
limit suggest that this TDE is closely related to the onset of a relativistic jet from a
supermassive black hole (hereafter BH). The jet is expected to be magnetically dominated \cite{bkg11}. Lei \& Zhang (2011)  suggested that Blandford-Znajek process (hereafter BZ, \cite{bz77}) is the plausible mechanism to launch the relativistic jet from this source \cite{bz77, lwm05, lwzz08}.They found
that the BH of this source carries a moderate to high spin \cite{lz11}. In addition, Lei, Zhang \& Gao (2013) 
interpreted a 2.7-day quasi-periodic variation with noticeable narrow dips in the X-ray light curve
by invoking a precessing, BZ powered jet collimated by a wind launched from a twisted and warped disk \cite{lzg13}.

The radio observation extending to about 600 days revealed unexpected features: about 1 month later, the radio emission showed a surprising re-brightening feature. Berger et al. (2012) and Zauderer et al. (2013) suggested that the radio re-brightening is a result of late-time energy injection from the central engine \cite{bzp12, zbm13}. However, to explain the late-time radio re-brightening, these models require that the energy of the source increases by a factor of 10-20 from 5 to 200 days, and there is no indication for this additional energy injection from X-ray observations. 

We show that this re-brightening can be naturally accounted for by a two-component jet model. The early radio emission is dominated by the inner (faster) jet, while the later radio observations can be interpreted by the outer (slower) jet. Besides, this model can also explain the two quasi-period variation features in X-rays lightcurve.

\section{Two-Component Jet Model for the Radio Light Curves and Spectra}
Sw J1644+57 was accompanied by bright radio emission. The radio observations extending to $t \simeq 26$ days were first presented in \cite{zbs11}. Later Berger et al. (2012) presented the data extending to $t \simeq 216$ days \cite{bzp12}, and Zauderer et al. (2013) presented the data extending to $t \simeq 600$ days \cite{zbm13}.

The millimeter flux on a timescale of $\sim 100$ days is significantly brighter than what is expected by extrapolating the early declining light curve based on a single jet model \cite{mgm12, bzp12}. This re-brightening can be naturally accounted for by a two-component jet model. We will show that our model can interpret the long-term radio data up to $\sim 600$ days.

Inspecting the observed X-ray light curve of Sw J1644+57, the jet power may evolve with time as $\propto t^{-5/3}$ at later times. In such case, the late time dynamics of jet just depends on the total ejected kinetic isotropic-equivalent energy $E_\mathrm{k,iso}$, which is about the energy injected in the initial emission episode \cite{zm01}.

We develop a numerical two-component model to fit the data. The dynamical evolution of the jets are followed using a set of hydrodynamical equations \cite{hgdl00}. Synchrotron spectra of both jets are calculated using the standard broken-power-law spectral model. We consider the two-component jet propagrates into a gaseous circumnuclear medium (CNM) with a number density $n$.
For a collimated jet, the jet effect becomes important when $1/\Gamma > \theta_\mathrm{j}$ \cite{zm04}. In the calculations, we include a suppression of the flux density due to this jet break effect \cite{zm04}. The steep spectral slope at lower frequencies indicate the existence of self-absorption \cite{zbs11, bzp12}. We therefore calculate the self-absorption frequency in detail.  

\begin{figure}[htbp]
\center
\includegraphics[width=140mm]{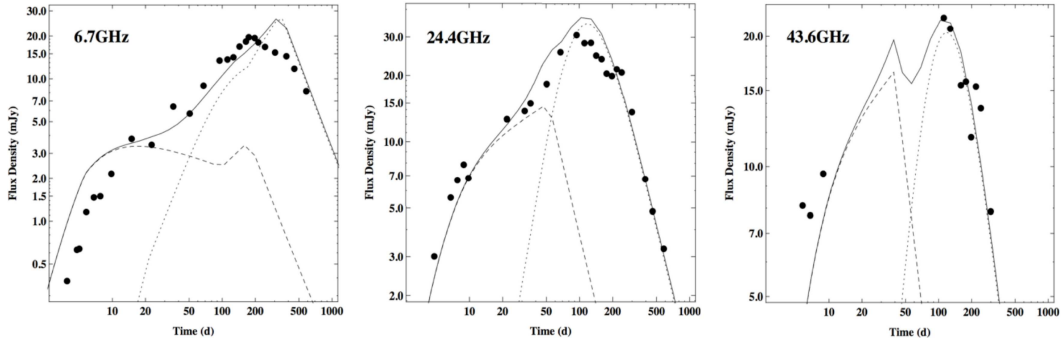}
\caption{Radio lightcurves (filled circles) of Sw J1644+57 extending to $t \simeq 600$ days. The solid lines are the model lightcurves using the two-component jet model. The details of model parameters are shown in Table 1. The earlier peak is mainly contributed by the faster inner jet (dashed lines), while the later peak is mainly contributed by the slower outer jet (dotted lines).}
\label{fig_lc}
\end{figure}

\begin{figure}[htbp]
\center
\includegraphics[width=140mm]{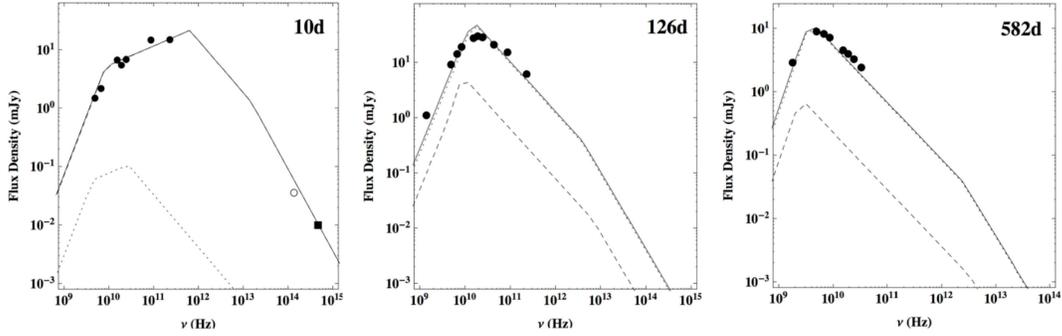}
\caption{Multi-frequency radio (filled circles) spectral distributions of Sw J1644+57 at $t \simeq 10, 126, 582$ days. The K/K$_s$ and R band flux at $t \simeq 10$ days are also shown with open circles and filled squares, respectively. The radio data are from \cite{zbm13}, and IR/optical data are from \cite{bkg11} and \cite{ltc11}. The solid lines are model fits based on our two-component jet model with the same parameters used in Fig. 1. The contributions from the inner and outer jets are shown by the dashed and dotted lines, respectively.}
\label{fig_spec}
\end{figure}

We apply the model to simulately fit the light curves as presented in Fig.\ref{fig_lc}, and the broad-band spectra as presented in Fig.\ref{fig_spec}. The best-fit parameters are summarized in Table 1.

\begin{table*}[htp]
\begin{center}
\caption{Results of radio observations fits \label{tb1}}
\begin{tabular}{ccccccccc}
\hline\noalign{\smallskip}
\hline\noalign{\smallskip}
    Jet component & n (${\rm cm}^{-3}$) & $\theta_{\rm obs}$ (deg) & $E_{52}$ & $\Gamma_{\rm j}$ & $\theta_{\rm j}$ (deg) & $p$ & $\epsilon_{\rm B}$ & $\epsilon_{\rm e}$ \\
\hline\noalign{\smallskip}
    inner-narrow & 0.25 & 7.0 &3.0   & 5.5 & 6.0 & 2.8 & 0.25 & 0.2 \\
    outer-wide & 0.25 & 7.0 &30.0   & 2.5 & 10.0 & 2.8 & 0.13 & 0.15 \\
\noalign{\smallskip}\hline
\end{tabular}
\end{center}
\end{table*}

\section{Possible Origin of the Two-component Jet}
The two-component jet model has been widely adopted to interpret data of active galactic nuclei (AGNs) and gamma-ray bursts (GRBs). For example, the limb-brightened morphology shown in several radio galaxies can be regarded as the evidence of a slower sheath-jet surrounding a faster spine-jet \cite{ggf04}; The rebrightening of XRF 030723 and the chromatic behavior of the broadband afterglow of GRB 080319B have been considered as evidence for a two-component jet in GRBs \cite{hwd04}. 

Xie et al. (2012) proposed a two-component jet model for both GRBs and AGNs. The baryon-poor inner jet is driven by the BZ mechanism \cite{bz77, lzl13}, in which the rotational energy of a BH is extracted to power the jet in the form of a Poynting flux via the open field lines penetrating the event horizon. A baryon-rich outer/wide jet can be launched centrifugally via the open magnetic field lines threading through the disk by the Blandford-Payne (BP)  process \cite{bp82}.

\begin{figure}[htbp]
\center
\includegraphics[width=60mm]{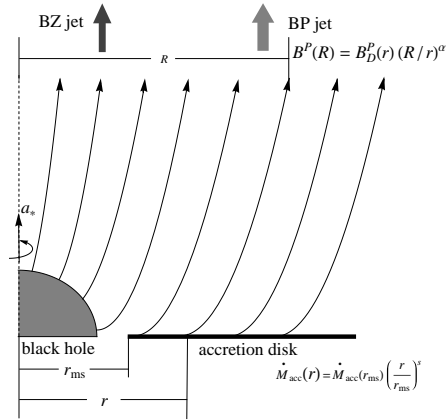}
    \caption{A schematic drawing of the magnetic field configuration for the two-component jet model, in which the inner-narrow-fast jet and the outer-wide-slow jet are driven by the BZ and BP processes, respectively \cite{xlz12}.}
    \label{Fig:fig1}
\end{figure}

For Sw J1644+57, the jet is expected to be magnetically dominated \cite{lz11}. The BZ and BP mechanisms play important roles in jet launching and the 2.7-d quasi-periodic variation in X-ray lightcurve \cite{lz11, lzg13}. We thus speculate that the two-component jet model may be powered by the BZ and BP mechanisms. The schematic picture of the model is shown in Fig.~\ref{Fig:fig1}. The the BZ process launches the inner jet via the open magnetic field threading the BH, while the BP process produces the outer jet via the open magnetic field threading the disk. 

\section{Interpretation of the X-ray Observations}
It is noted that two quasi-periodic variations of Swift J1644+57 were detected after a few
days since the BAT trigger \cite{rmr12, bkg11, sswk12}.
The first is a rough 2.7-day periodicity in the dips of the XRT light curve of Sw J1644+57. This quasi-periodic signal was first pointed out by \cite{bkg11}, more carefully studied by \cite{sswk12}, and then further confirmed by \cite{lzg13} using the stepwise filter correlation (SFC) method \cite{gzz12}. The second QPO feature was discovered by \cite{rmr12}, who produced the light curves of Sw J1644+57 over the 0.2-to-10.0-keV energy band using all the observational data from different detectors. The 2-to-10-keV power density spectra of both the Suzaku and first XMM-Newton observations displayed a potential QPO component near 5 mHz.

These two apparent quasi-periodic features are further evidence for the two-component jet model: 
The 200s QPO is related to the inner jet launched from the horizon of a spinning BH
surrounded by a warped accretion disk, and the 2.7-days quasi-periodic variation is associated with
the outer processing jet launched near the Bardeen-Petterson radius.

An advantage of this model is that it may lead to efficient magnetic dissipation.  
Since the inner (BZ) and outer (BP) jets move relatively with each other through precession and differential motion due to different Lorentz factors, dissipations of magnetic energy in both jets are inevitable. One can speculate two types of interactions. One is ``collision'' between the inner jet and the outer jet as the former streams into the later, the other is the relative ``shearing'' at the boundary of two jets. These would induce significant magnetic dissipation, through processes similar to the ICMART process envisaged in GRBs \cite{zy11} or something similar to Kelvin-Helmholtz instability (in the strongly magnetized regime).  Using the X-ray observations and the radio data fits, we can estimate the magnetic dissipation efficiency as $\eta =  E_{\rm X, iso} /E_{\rm j, iso} \simeq 0.4$.  Such an efficiency is quite high, comparable to that of GRBs, and consistent with the magnetic jet dissipation models \cite{zy11}.

\section{Conclusions}
The two-component jet scenario naturally predicts two afterglow components, which nicely interpret the radio data of the source, especially the mysterious re-brightening $\sim 100$ days after the trigger. Such a model is also motivated by two natural jet-launching mechanisms, i.e., the BZ process for powering the jet from the BH horizon and the BP process for driving an outflow from the warped disk at Bardeen-Petterson radius
\cite{bp82, lzg13}. In addition, the two quasi-periodic variations (200s and 2.7-day) in the X-ray emission are also well explained by this model.

\end{document}